\documentstyle[a4,12pt]{article}

\newcommand{\beq}{\begin{equation}}
\newcommand{\eeq}{\end{equation}}
\newcommand{\ba}{\begin{eqnarray}}
\newcommand{\ea}{\end{eqnarray}}
\newcommand{\bi}{\begin{itemize}}
\newcommand{\ei}{\end{itemize}}

\def\ps@headings{}
\def\@chapapp{}
\begin{document}
\begin{titlepage}
\vspace{0.2in} 
\begin{flushright}
MITH-98/10  \\ 
\end{flushright} 
\vspace*{1.5cm}
\begin{center} {\LARGE \bf A tale of two condensates:\\
the odd "Bose - Einstein" condensation of atomic Hydrogen.\\} 
\vspace*{0.8cm}
{\bf  E.~Del Giudice} and {\bf  G.~Preparata} \\
{\it Dept. Physics and INFN-via Celoria, 16-20133 MILAN  \\ 
ITALY }
\\ \vspace*{1cm}
\end{center} 
\begin {abstract}

The recent report of the observation of Bose-Einstein condensation in atomic
Hydrogen,  characterized by an "anomalous" density spectrum, is shown to be in
agreement with the prediction of the  existence of two condensates for 
temperatures lower than a well defined temperature (which for  Hydrogen is $
105~ \mu K $ ), based on the QED coherent interaction in a gas of ultracold
atoms at a density $n > n_{0}$  ($n_{0}=\left( \frac{1}{\lambda}\right)^{3}$, 
$\lambda$  being the wave-length of the e.m. modes resonantly coupled to the
Hydrogen atoms).

\end{abstract}

\baselineskip=12pt 
\vfill \begin{flushleft}  Milan, 15 November 1998 \\
%PACS 33.10, 62.50, 64.70  \vspace*{3cm} \\
%\noindent{\rule[-.3cm]{5cm}{.02cm}} \\
%\vspace*{0.2cm} \hspace*{0.5cm} ${}^{a)}$ 
%E-mail address: xue@mi.infn.it
\end{flushleft} 
\end{titlepage}
\baselineskip=12pt

The last few years have seen a remarkable surge of activity in the physics of
ultracold atoms that  has led to the discovery of a phase transition from the
gaseous state to a new state, characterized by a  very narrow momentum
distribution and (relatively) high density.  By combining definite improvements 
in the techniques of magnetic trapping and r.f. evaporative cooling in 1995,
within a few months, three  independent groups were able to show such
transition in the dilute alkali gases of $Rb$ \cite{1}, $Na$ \cite{2} and $Li$
\cite{3}.  Only a
few months ago another group \cite{4}, through a remarkable experimental feat,
was able to  detect such transition in atomic Hydrogen, thus confirming the
(apparent) universality of its nature. 

An exciting aspect of such discoveries
is that they appear to finally give, after 70 years of  waiting, the
confirmation of a fundamental prediction that, in the framework of the
revolutionary  quantum physics, S. N. Bose \cite{5} and A. Einstein \cite{6}
were able to
derive from the general ideas of the  statistical thermodynamics of Maxwell and
Boltzmann.  According to Bose and Einstein, in order for the  perfect gas to
obey the "heat theorem" of W. Nernst \cite{7} (better known as the third
principle of  thermodynamics) above a certain (number) density, the
Bose-Einstein  density (we use throughout this  paper the natural units system,
where $\hbar=c=k_{B}=1$ )
\ba
n_{BE}=2.612 \left(\frac{mT}{2 \pi} \right)^{\frac{3}{2}} ,
\label{eq1}
\ea
where $m$ is the mass of the atoms, a number of atoms ($V$ is the volume of
the gas)
\ba
 N_{c}=(n-n_{BE}) V 
\ea 
leaves the chaotic world of the gaseous state to populate the ground state with
momentum $|\vec{p}|=0$. In  this way Bose and Einstein finally showed how
nature manages to achieve in a continuous manner zero  entropy at $T=0$, as
demonstrated by Nernst in a countless number of physical systems \cite{7}, in 
disagreement with the highly singular behavior predicted by classical physics.

The fact that the transition (apart from the notable exception of Ref.\cite{1})
occurs
at the predicted  value 
%~\ref{eq1}
(1), and its most remarkable signal is the dramatic
narrowing of the momentum and space  distributions of a  number of atoms, in agreement with
the predictions based on the structure of the magnetic trap, left no  room for
doubting
that Bose-Einstein condensation finally belongs to the realm of
natural phenomena.   The equally dramatic discovery, a year later, of stunning
interference patterns, with the typical de- Broglian modulation \cite{8}, has 
begun,
however, to raise doubts in some of us that the observed  condensation could in
fact be exactly what was predicted by Bose and Einstein. The problem is an old
one and in the case of superfluidity and superconductivity has been  intensely
argued by leading theoretical physicists \cite{9}: it boils down to the simple
question: can a Bose- Einstein condensate (BEC) exhibit a macroscopic phase?  A
straightforward analysis of the Quantum  Field Theory (QFT) of the statistical
thermodynamics of a perfect gas (which the observed dilute atomic  systems
approximate very accurately) unambiguously shows that the BEC cannot posses a
well defined  phase for lack of "sufficient reason".   Indeed the
non-interacting nature of the condensed atoms  prevents them from developing the
peculiar phase relations that characterize a coherent state.  The situation 
might change if some interaction is introduced, but, as we have argued in Ref.
10, it appears impossible  to generate a phase (and a robust one, as
experimentally shown in a subsequent experiment \cite{11}) with  the short range
interactions that are available in the generally accepted approach to condensed
matter.   Thus the surprising results of Ref. \cite{8} and Ref. \cite{11}, in 
the light of
the essentially negative answer to the  question whether a BEC may acquire a
macroscopic phase, appear to exclude that the condensates  observed in Refs.
\cite{1},\cite{2} and \cite{3} are BEC's.  If not BEC's what else can they be? 
In a recent paper
\cite{10} we have examined the problem in the framework of QED, and of its 
unexpected and neglected coherent long-range interactions \cite{12}. We have 
shown
that when the density  $n_{c}$ of the atomic systems in the condensate is larger than
\ba                                                 
n_{0}=\left(\frac{1}{\lambda} \right)^{3}, 
\ea
where $\lambda$ is the wave-length of the e.m. transition between the ground
state and the first excited state, a  completely new type of condensation may
occur, driven by the energy gain that the system can achieve  by letting its
atoms oscillate in phase with resonant modes (of wave-length $\lambda$) of the e.m.
field.  Let's now  summarize the main results of the analysis of Ref. \cite{10}:
\begin{description}
\item [{\rm i)}] in the new condensed phase, which we shall call Coherent Electrodynamic Condensate (CEC), 
an average energy gap $\bar{\delta}$  develops, whose value is given by
\ba
\bar{\delta}=\frac{3}{16 \pi^{2}}\frac{\omega^{2}}{m_{e}}f x_{c}=
\bar{\delta_{0}} x_{c} 
\ea
where $\omega=\frac{2\pi}{\lambda}$  is the energy of the atomic transition,
$ f$ its oscillator strength and $x_{c} (0 \le x_{c} \le 1)$  
the fraction of atoms that belong to the CEC;
\item [{\rm ii)}] the fraction $x_{c}$ is given by the solution of the following 
equation
\ba
1-x_{c}=\frac{1}{n_{c}}\left(\frac{mT}{2 \pi} \right)^{\frac{3}{2}} 
f \left( \frac{\bar{\delta_{0}} x_{c} }{T} \right),
\ea
where
\ba
f(x)=\frac{2}{\sqrt{\pi}} \int_{0}^{\infty}\frac{dt t^{\frac{1}{2}}}
{(e^{t+x}-1)}.
\ea
By use of Eq. ~\ref{eq1} we can cast (5) in the form 
\ba
\frac{1-x_{c}}{x_{c}}  =\frac{n_{BE}}{n_{0}} 
\frac{f \left( \frac{\bar{\delta_{0}} x_{c} }{T} \right)  }{2.612} ;
\ea
\item [{\rm iii)}]
when $n_{BE} < n_{0}$, and this happens when $T<T_{BEC}$,  where
\ba
T_{BEC}=\frac{2 \pi}{m} \left( \frac{n_{0}}{2.612}\right)^{\frac{2}{3}},
\ea
there is a density interval
\ba
 n_{BE}<n<n_{0}
\ea
where our theory predicts an incoherent BEC, in full agreement with the theory
of Bose and  Einstein.
\end{description}
In table I we report the relevant parameters of our theory for$ H$, $Li$, $Na$
and $Rb$ respectively,  while in Figs. 1, we show the phase plane $T-n$ for the
different atomic species.  Noting that in the actual experiments $T_{Na}=2~ \mu
K$  \cite{2}, and $T_{Rb}= 140~nK$ \cite{3}, are all above their respective
$T_{BEC}$, our theory predicts that what has been  observed in those
experiments is indeed not a BEC but a CEC, and the observed macroscopic phases
are  in complete agreement with our
predictions. 

Let's now focus our attention upon the very recent report \cite{4} of
condensation in atomic Hydrogen,  where the temperature at which the transition
occurs is $T=44~\mu K$, definitely lower than $T_{BEC}=105 ~\mu K$.Thus our 
theory predicts the existence of two condensates: the BEC for
\ba
1.48~10^{14}~cm^{-3} < n <5.57~10^{14}~cm^{-3} ,
\ea
and the CEC for
\ba
n>5.57~10^{14}~cm^{-3}.
\ea
Is there any evidence for this prediction in the reported data?  Before we
answer this question, let us try  to follow the steps of the transition to the
condensed phase(s).  When lowering the temperature from  $T=53~\mu K$ to 
$T=44~ \mu K$ the
MIT group finds an extremely suggestive change in their spectra of the two-
photon $1S-2S$ transitions, signalling that a number $N_{c}$ of the initial atoms
have made a transition to the  BEC.  In a time determined by the typical trap
parameters (milliseconds) the system reaches an  equilibrium distribution
[4]
(V is the profile of the magnetic trap)
\ba
n=n_{p}-\frac{V(\rho,z)}{\tilde{U}}
\ea
where the peak density $n_{p}$ is related to the (unknown) number $N_{c}$ by the
formula [4]
\ba
N_{c}=\frac{16 \pi \sqrt{z}}{15}
\frac{{\tilde{U}}^{\frac{3}{2}}n_{p}^{\frac{5}{2}}}{\omega_{\rho}^{2}
\omega_{z}m^{\frac{3}{2}}},
\ea
where $\omega_{p}$ and $\omega_{z}$ are the radial and longitudinal trap
"pulsations" respectively, and $\tilde{U}=\frac{4 \pi a}{m}$  describes the 
short-range repulsion between the cold atoms.   From the experimental 
determination \cite {4}
\ba
n_{p}=(4.8 \pm 1.1)~10^{15}~cm^{-3},
\ea
Eq. (13) leads to the evaluation
\ba
N_{c}=(1.1 \pm 0.6)~10^{9}.
\ea
Now, only if $n_{p}>n_{0}=5.57~10^{14}~cm^{-3}$ the CEC  will after a while 
\footnote{ Even though at present we have not carried out this analysis, we
expect that it will take a relatively long time   before the CEC phase emerges 
from an incoherent system of ordered atoms (BEC). }  develop.  Due  to the
average gain of $\bar{\delta_{0}} \simeq 100~\mu K $ per atom, in fact, that
portion of the original BEC that lies  between $n_{p}$ and $n_{0}$ will fall in
the CEC, leaving the rest, for which $n < n_{0}$ in the BEC phase. The
experimental determination (14) shows that  we are indeed in the situation
where two distinct  condensates coexist around the bottom of the magnetic trap. 
Recalling that in the frequency shift  analysis of the densities of ref.
\cite{4} $n_{p}$ corresponds to $ \Delta \nu_{p}=-(9. \pm .15)~MHz$, $n_{0}$ 
to  $\Delta \nu_{0}=-(106. \pm 20) ~kHz$, while $n_{BE}$ corresponds to $\Delta
\nu_{BE}=-(28. \pm 6) ~kHz$, the observed  Doppler-free spectrum (Figs 2 and 3
of Ref.~\cite{4}) can be easily understood.  Indeed the CEC matches  perfectly
the distribution of Fig. 2, including its stopping at about $150~ kHz$ ($\Delta
\nu_{0}=-(106. \pm 20) ~kHz$), the (almost) empty region in Fig. 3 between
$-100$ and $-80~ kHz$ can be understood by the fact that the  gas fluctuations
of  typical density $n_{BE}$ ( $\Delta \nu_{BE}=-30.  ~kHz$~) will cause the
BEC condensate at densities  within $n_{BE}$ from $n_{0}$ to fall sooner or
later in the "deeper  trap" (by $ \sim 100 \mu K$) of the CEC condensate.   
Finally the gas fluctuations around $n=n_{BE}$ will tend to replenish this
region of the BEC condensate.  The  net consequence of this (semiquantitative)
discussion is that we can now explain {  \it  "~...spectral weight at 
frequency shifts much larger than expected for the maximum density in the
normal gas.  The origin of  this effect is (not yet) understood \/ "} \cite{4}. 
And we hope that {\em our} parentheses will be reckoned by the  reader as an
adequate, necessary modification of the discussion of the Authors of Ref.
\cite{4}. 

We hope that the natural, quantitative explanation of the odd-looking results on
the condensation of  cold atomic Hydrogen \cite{4} achieved by our theory,
rigorously based on QED, will contribute to open the  eyes of the condensed
matter physics community to an approach to the subject which, without 
violating any fundamental laws of physics, enriches the field with an
extremely wide range of new  interaction mechanisms, whose real potential for
new, exciting developments remains still largely  undisclosed.

\newpage
\begin{center}
{\Large  TABLE I}\\
\end{center}
\begin{center}
\begin{tabular}{|c|c|c||c||c||c||c||c|} \hline
Atom & $\lambda(A)$ & $ \omega (eV)$ &$f$ &$n_{0} ~cm^{-3}$ &
$\delta_{0}(\mu K)$ & $\nu_{BE} ~\frac {cm^{-3}}{(\mu K)^{\frac{3}{2}}}$ &
$T_{BEC} (\mu K)$  \\ \hline
H  & $ 1215.67$ &  $10.34 $ & $.44  $ & $5.57~ 10^{14} $ & $200  $ & 
$5.08~10^{11}$ & $105$ \\ \hline
Na & $ 5889.9 $ &  $2.13 $ & $.64  $ & $4.89~ 10^{12} $ & $18.7  $ & 
$5.62~10^{13}$ & $0.20$ \\
~ & $  5895.5 $ &  $ 2.13 $ & $ .32 $ & $ 4.87~ 10^{12} $ & $9.35 $ & ~ & ~
\\ \hline
Li & $ 6707   $ &  $1.87 $ & $.50  $ & $3.31~ 10^{12} $ & $11.25 $ & 
$9.40~10^{12}$ & $0.49$  \\ 
~ & $  6707   $ &  $ 1.87 $ & $ .25 $ & $ 3.31~ 10^{12} $ & $5.623$ & ~ & ~
\\ \hline
Rb & $ 7800  $ &  $1.61 $ & $.67  $ & $2.11~ 10^{12} $ & $11.17 $ & 
$4.02~10^{14}$ & $0.03$ \\ 
~ & $  7947   $ &  $ 1.58 $ & $ .33 $ & $ 2.12~ 10^{12} $ & $5.58 $ & ~ & ~
\\ \hline 
\end{tabular}
\end{center}
~\\
{ \small The CEC parameters for different atomic species.}
\newpage
\begin{center}
FIGURE CAPTION
\end{center}
\begin{figure}[h]
\small
\caption{  The phase diagram for the condensation of H Atoms. The
lines  are the boundaries of the regions where
the Coherent Electrodynamics Condensation (CEC) occurs.}
\end{figure}

\end{document}